\pdfoutput=1
\documentclass[11pt]{article}

\usepackage{graphicx}
\usepackage{amsmath}
\usepackage{amssymb}
\usepackage{dcolumn}
\usepackage{bm}
\usepackage{slashed}
\usepackage{color}
\usepackage{authblk}
\usepackage{ulem}
\usepackage{CJK}

\def\L{{\Lambda}}
\def\d{{\delta}}

\def\e{{\epsilon}}

\def\a{{\alpha}}
\def\b{{\beta}}

\def\g{{\gamma}}
\def\G{{\Gamma}}

\def\h{\eta}
\def\p{{\pi}}

\def\m{{\mu}}
\def\n{{\nu}}
\def\r{{\rho}}
\def\s{{\sigma}}
\def\S{{\Sigma}}

\def\x{{\xi}}

\def\({\left(}
\def\){\right)}
\def\[{\left[}
\def\]{\right]}

\newcommand{\pd}{{\partial}}
\newcommand{\dg}{\dagger}

\newcommand{\tr}{\text{tr}}

\date{\today}

\begin{document}

\begin{CJK}{UTF8}{gbsn}

\title{\bf In-medium Electromagnetic Form Factors and Spin Polarizations}

\author[1]{{Shu Lin}
\thanks{linshu8@mail.sysu.edu.cn}}
\affil[1]{School of Physics and Astronomy, Sun Yat-Sen University, Zhuhai 519082, China}
\author[1]{{Jiayuan Tian}
\thanks{tianjy26@mail2.sysu.edu.cn}}

\maketitle


\begin{abstract}

Despite significant progress having been made in understanding spin polarization phenomena in relativistic heavy ion collisions, the underlying assumption of weakly coupled medium is not well justified. To go beyond this limitation, we formulate spin polarization in background electromagnetic fields using form factors. We show that the form factors at tree level in vacuum reproduce the spin polarization effects found in chiral kinetic theory. The vacuum form factors corresponding to spin couplings to perpendicular electric fields, parallel and perpendicular magnetic fields are degenerate. The degeneracy can be lifted in medium. As an example, we calculate the in-medium QCD radiative corrections to the form factors at one-loop order, where we find partial lift of the degeneracy: the spin couplings to parallel and perpendicular magnetic fields are different, but the spin couplings to perpendicular electric and parallel magnetic fields remain the same. 
\end{abstract}

\newpage

\section{Introduction}

The spin physics of relativistic heavy ion collisions (HIC) has attracted much attention over the past two decades. Based on spin-orbital coupling, early theories predicted that the final particles in off-central heavy ion collisions should be spin polarized \cite{Liang:2004ph,Liang:2004xn}. Indeed, spin polarization has been observed for $\L$ hyperon in low energy collisions \cite{STAR:2017ckg,STAR:2019erd}. Many theoretical efforts have been made in gaining quantitative understanding of the $\L$ hyperon spin polarization from spin-vorticity coupling \cite{Becattini:2013fla, Fang:2016vpj, Li:2017slc, Liu:2019krs, Becattini:2017gcx, Wei:2018zfb, Wu:2019eyi, Fu:2020oxj, Zhang:2019xya, Weickgenannt:2020aaf, Gao:2012ix} and spin-shear coupling \cite{Hidaka:2017auj, Liu:2021uhn, Becattini:2021suc, Fu:2021pok, Becattini:2021iol, Yi:2021ryh, Lin:2022tma}. Spin also couples with electromagnetic fields besides fluid velocity gradient, which are known as spin-magnetic coupling (SBC) and spin Hall effect (SHE) \cite{Liu:2020dxg, Mameda:2022ojk}. The latter has been under intensive investigations in condensed matter systems \cite{RevModPhys.87.1213, PhysRevLett.95.226801}. Despite short lifetime of electromagnetic fields in HIC, effective electric field can also be realized with baryonic density gradient, making SHE potentially observable \cite{Fu:2022oup}.

The interaction of fermions with external electromagnetic (EM) fields is interesting on its own. In the weak field regime, the spin interaction can be systematically described by spin kinetic theory \cite{Gao:2019znl, Weickgenannt:2019dks, Hattori:2019ahi, Liu:2020flb, Sheng:2020oqs,Yang:2020hri,Weickgenannt:2021cuo,Sheng:2021kfc,Lin:2021mvw}, which is a massive generalization of chiral kinetic theory (CKT), see \cite{Hidaka:2022dmn} for a recent review. In the strong magnetic field regime, the spin interaction can be described by kinetic theory with Landau level basis \cite{Sheng:2017lfu, Lin:2019fqo, Hattori:2016lqx, Fukushima:2019ugr}. The strong magnetic field is also known to have interesting effect in thermodynamics and transport properties \cite{Zhang:2020ben,Fang:2021ndj,Fukushima:2015wck,Hattori:2016lqx,Fukushima:2017lvb,Hattori:2017qih,Hattori:2016njk,Lin:2021sjw,Peng:2023rjj}, see \cite{Hattori:2023egw} for a review and references therein. 

Our present knowledge of spin polarization in background EM fields mostly relies on the assumption that particles interact weakly with medium, which is however not well justified for quark-gluon plasma (QGP) produced in HIC. On the other hand, we know from field theory that the SBC adopts a description a la fermion scattering on background EM fields. A symmetry-based parameterization of the interaction vertex using EM form factors allows us to study the SBC with more general couplings. Indeed, the scattering picture deals with the same on-shell fermions as in the CKT framework. In this work, we make the connection between scattering picture and CKT framework more precise: the gradient expansion in CKT corresponds to the limit of low momentum transfer in the scattering picture. We first reproduce the CKT results of spin polarization by a field theory calculation of correlation functions in background EM fields. It is manifest in the calculation that the interaction vertex is the vacuum one with Lorentz symmetry.


We then give general parameterization of the vertex using form factors.
Electromagnetic and gravitational form factors in vacuum have been extensively studied in different branches of physics, cf \cite{Giunti:2014ixa,Polyakov:2018zvc}. The main difference in our case is the presence of medium, which breaks the Lorentz symmetry. While in vacuum the SBC and SHE has degenerate couplings, in the medium, they split into three non-degenerate couplings in general: SHE, spin parallel and perpendicular magnetic couplings.

As an example, We evaluate in-medium EM form factors for quarks in QGP\footnote{The in-medium gravitational form factors have been discussed and radiative corrections have been considered for spin-vorticity coupling in \cite{Buzzegoli:2021jeh,Lin:2023ass}.}. Since current quark mass is much smaller than temperature of QGP, we will ignore quark mass and work with Weyl fermions. We find the in-medium form factors in this example lift the degeneracy of spin coupling with parallel and perpendicular magnetic fields. We also discuss its implication for phenomenology of spin polarization in HIC.




The paper is organized as follows: in Sec.~\ref{sec_2}, we perform field theory calculations of fermion correlation function in background EM fields at tree level in medium, which establishes equivalence with CKT results and reveals the Lorentz invariant nature of the coupling with EM fields; in Sec.~\ref{sec_3}, we identify a set of kinematic conditions that enables us to interpret the correlation function as a scattering process, which can be described by form factors. We first define three degenerate vacuum EM form factors corresponding to spin couplings to perpendicular electric, parallel and perpendicular magnetic fields. We then generalize to in-medium EM form factors with lifted degeneracy in general; in Sec.~\ref{sec_4}, we calculate the in-medium QCD radiative corrections to EM form factors as an example, where we find partial lift of the degeneracy; Sec.~\ref{sec_outlook} is denoted to summary and outlook. Detail of calculations are reserved in appendices.

\section{Tree level fermion correlation function in EM fields}\label{sec_2}

In this section, we consider the fermion correlation function in weak external EM fields. 
We consider the Wigner functions for fermions defined as follows:
\begin{align}
    S^{<(>)}(X=\frac{x_1+x_2}{2}, K)= \int d^4(x_1-x_2) e^{i K \cdot (x_1-x_2)}S^{<(>)}(x_1,x_2)U(x_2,x_1),
\end{align}
with the weak field condition given by $\pd_X\ll K$, i.e. the gradient from the EM fields much less than particle momentum. The EM fields enter as background fields in the lesser(greater) propagators $S_{\a\b}^{<}(x_1,x_2)=\langle \psi^{\dagger} _\b(x_2)\psi_\a(x_1) \rangle$ and $S_{\a\b}^{>}(x_1,x_2)=\langle \psi_\a(x_1)\psi^{\dagger}_\b (x_2) \rangle$, as well as through the gauge link $U(x_2,x_1)=e^{-i\int_{x_1}^{x_2} dz^\m A_\m(z)}$ introduced to maintain gauge invariance\footnote{We have absorbed the electromagnetic coupling constant $e$ in the definition of EM fields.}. This quantity is extensively studied in CKT by solving from Dirac equation in external EM fields \cite{Hidaka:2016yjf, Hidaka:2017auj}. For right-handed fermions, chiral components of lesser propagator $S^{<\mu}$ up to first order in gradient expansion reads:
\begin{align}\label{Sless_CKT}
    S^{<\mu }(X, K)=\frac{1}{2}\tr[\s^\m S^<(X, K)] = -2\pi \left[\delta (K^2)K^\mu \tilde{f}(k_0) + \frac{1}{2}\epsilon^{\mu \nu \alpha \beta }K_\nu F_{\alpha \beta }\delta '(K^2) \tilde{f}(k_0) \right],
\end{align}
where $\sigma^\mu=(1,\boldsymbol{\sigma})$, and $\tilde{f}(k_0)$ is the Fermi-Dirac distribution function. The second term corresponds to the response to 
the EM fields. In particular, the spatial components read
\begin{align}\label{CKT_pol}
    \d S^{<i}=-2\p \big[k^i\d(K^2)+(k_0B^i-\e^{ijk}k_jE_k)\d'(K^2)\big]\tilde{f}(k_0),
\end{align}
containing both SBC and SHE. These effects are not specific to medium, as can be seen from the fact that the corresponding spectral function in the square bracket is independent of the medium. It shall be clear that the medium independence of the spectral function is tied to the vacuum type interaction of the fermions and EM fields. The medium independence ceases to hold when we consider medium modified interaction in the next section.

Now we turn to the field theory calculation of $S^<$ in an equilibrium medium. It can be calculated directly using the real time formalism, but the diagrammatic analysis is complicated because there are both type $1$ and $2$ vertices in Schwinger-Keldysh contour \cite{le2000thermal}. We will instead calculate $S^{ra}$ using the $ra$-basis, from which we can obtain $S^<$ through the Kubo-Martin-Schwinger (KMS) relation.
The spinor fields in $ra$-basis and $12$-basis are related as \cite{Chou:1984es}:
\begin{align}
    \psi _r = \frac{1}{2}(\psi _1 + \psi _2), \quad \psi _a = \psi _1 - \psi _2,
\end{align}
with the $r$ and $a$ fields interpreted as average and fluctuation fields respectively.
In the absence of the EM fields (indicated by the superscript $(0)$), the non-vanishing propagators in thermal equilibrium in the $ra$-basis are given by:
\begin{align}
    S_{rr}^{(0)}(K)&=\(\frac{1}{2}+\tilde{f}(k_0)\)2\p(K\cdot\bar{\s})\d(K^2),\label{S_prop1}\\
    S_{ra}^{(0)}(K)&=\frac{iK \cdot \bar{\sigma}}{K^2+i \eta\,\varepsilon(k_0)}, \label{S_prop2}\\
    S_{ar}^{(0)}(K)&=\frac{iK \cdot \bar{\sigma}}{K^2-i \eta\,\varepsilon(k_0)},\label{S_prop3}
\end{align}
where $\bar{\sigma}^\mu=(1,-\boldsymbol{\sigma})$, $\eta \to 0^+$. Without loss of generality, we focus on particles with positive energy thus $\varepsilon(k_0)=+1$. Using the KMS relation, the spectral function is given by:
\begin{align}
    \rho (K) = S_{ra}^{(0)}-S_{ar}^{(0)} = 2\pi (K \cdot \bar{\sigma})\delta (K^2) .\label{spectral_function}
\end{align}
The lesser correlation function can then be expressed as
\begin{align}
    S^<_{(0)}(K)=-\tilde{f}(k_0)\rho (K)=-2\pi (K \cdot \bar{\sigma})\delta (K^2)\tilde{f}(k_0) \label{free_lesser_propagator}.
\end{align}

\begin{figure}
  \centering
    \includegraphics[width=1\textwidth]{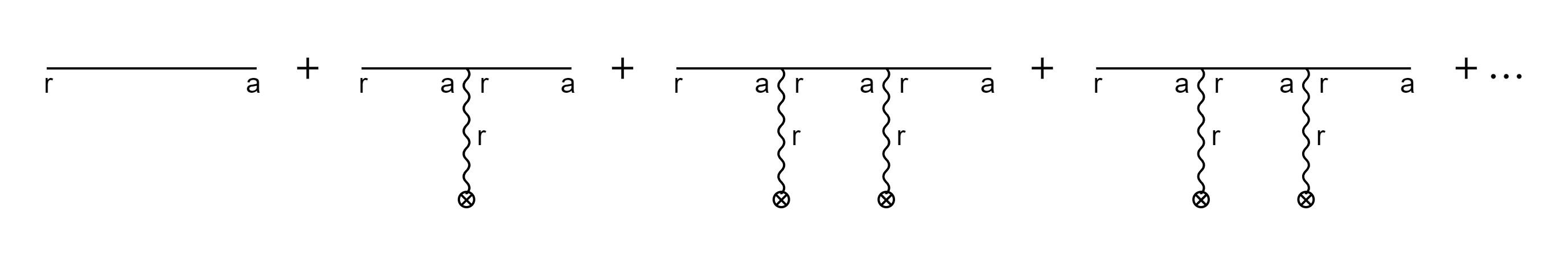}
  \caption{Retarded fermion correlation function with arbitrary insertions of external electromagnetic fields.}
   \label{tree_level_resum}
\end{figure}

Now we consider the case with EM fields. In the $ra$ basis, the external EM fields are naturally represented as the average field $A_r$. The correlation function in external EM fields can be represented by the sum of a series of Feynman diagrams as shown in Figure \ref{tree_level_resum}. Note that for $S^{ra}$ there is only one possible labelings of $ra$, which significantly simplifies the analysis.

Note also that there are momentum inflow at each vertex (to be denoted by $Q$). Since $Q\sim\pd_X$ characterizes the gradient, we shall equivalently perform expansion in $Q$ in the analysis below. The expansion arises either from the gauge link or from the interaction vertex.
The expansion of the gauge link is simple: taking a straight path for the gauge link as in the CKT approach, we obtain
\begin{align}\label{U_expansion}
U(x_2,x_1) = e^{-i\int^{x_2}_{x_1}dz \cdot A(z)} = e^{-iA (X)\cdot (x_2-x_1)}+O(\pd_X^2).
\end{align}
The correction at the second order $O(\pd_X^2)\sim O(Q^2)$ is beyond the order of our interest.
To perform the expansion in $Q$ from the vertex, we rewrite the Wigner function as
\begin{align}\label{Wigner}
S^{ra}(K) &= \int_s e^{iK \cdot s} S^{ra}(X,s)U(x_2,x_1) \nonumber\\
&= \int_s e^{iK \cdot s} \int_P e^{-iP \cdot s} S^{ra}_{re}(P)e^{iA \cdot s},
\end{align}
where $\int_s= \int d^4s$, $\int_P= \int \frac{d^4P}{(2\pi )^4}$, and $S^{ra}_{re}(P)$ is the correlation function resummed to all order in $A$ and up to first order in the momentum inflow $Q$. We split the contributions into zeroth and first orders in $Q$. The zeroth order contribution corresponds to resummation of constant $A$ field, which can be done conveniently as 
\begin{align}
S_{re}^{ra}(P)&=\frac{iP\cdot\bar{\s}}{P^2+i\h}+\frac{iP\cdot\bar{\s}}{P^2+i\h}(-iA\cdot\s)\frac{iP\cdot\bar{\s}}{P^2+i\h}+\cdots\nonumber\\
&=\frac{1}{iP\cdot\s}+\frac{1}{iP\cdot\s}(-iA\cdot\s)\frac{1}{iP\cdot\s}+\cdots\nonumber\\
&=\frac{1}{i(P-A)\cdot\s}=\frac{i(P-A)\cdot\bar{\s}}{(P-A)^2+i\h}.
\end{align}
Combining with the gauge link and performing the remaining integrals, we obtain
\begin{align}\label{O1}
S_{ra}(K)&=S_{re}^{ra}(K+A)=\frac{iK\cdot\bar{\s}}{K^2+i\h},\nonumber\\
&\Rightarrow\; S_{ra}^\m(K)=\frac{iK^\m}{K^2+i\h}.
\end{align}
In the above $P$ and $K$ can be identified as canonical and kinetic momenta respectively.

\begin{figure}
	\centering
	\includegraphics[width=.8\textwidth]{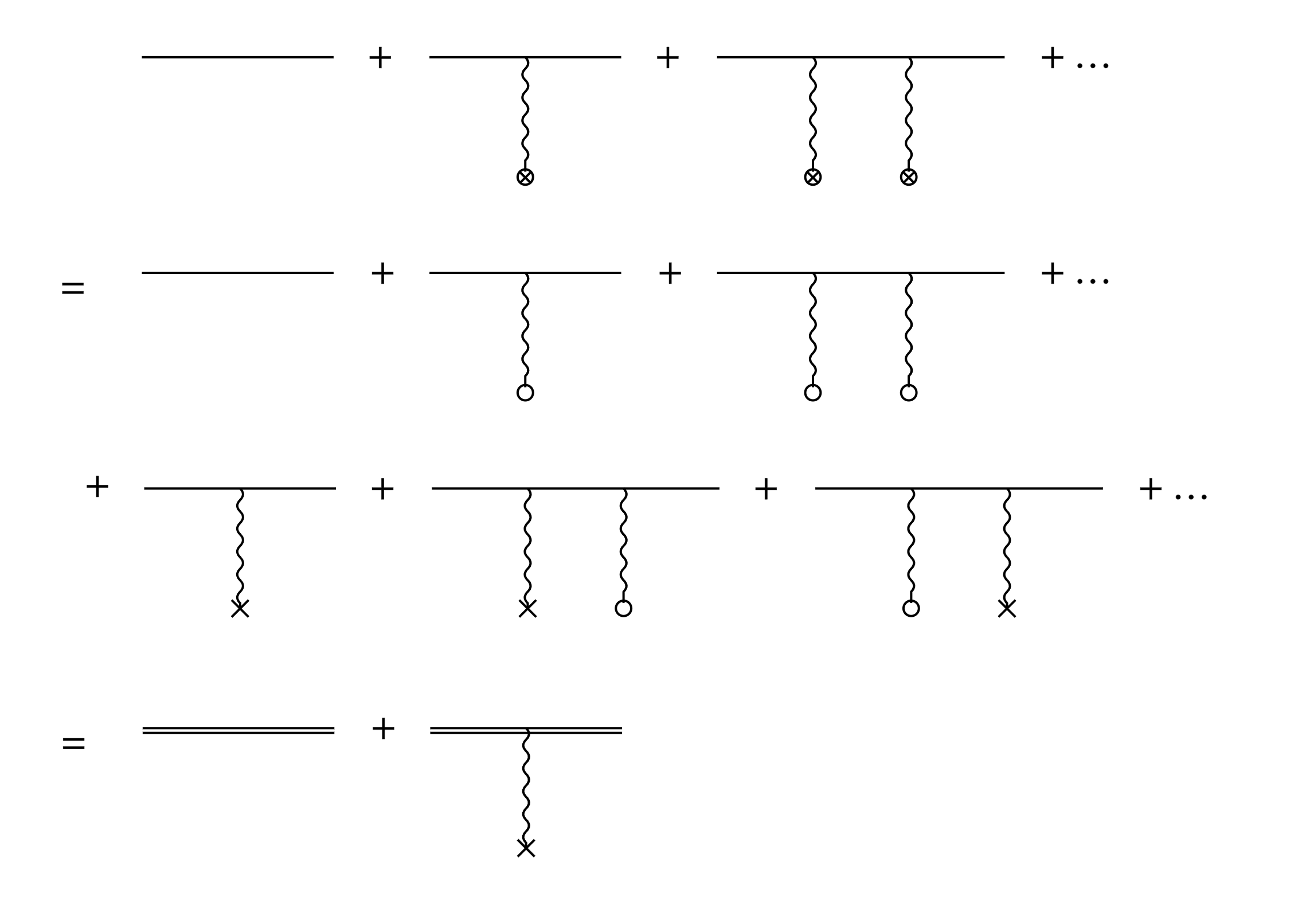}
	\caption{The diagrammatic contributions are split into $O(Q^0)$ and $O(Q)$ corresponding to second and third lines. The encircled crosses and empty circles denote EM fields with momentum inflow and without. The bare circles denote the difference of the two symbols, i.e. $O(Q)$ contribution from the EM fields. The double lines correspond to resummed correlation function in constant $A$ field.} \label{split}
\end{figure}

The result at the first order $O(Q)$ comes from diagrams with all constant $A$ except one carrying momentum $Q$, shown schematically in Figure~\ref{split}.
%
Using the diagram representation in Figure \ref{split}, we find it is given by the convolution of two resummed correlation functions and a vertex expanded to $O(Q)$. Note that the momentum inflow causes an $O(Q)$ order shift of momentum at the vertex. The momenta before and after the inflow can be found from the following Fourier factors
\begin{align}
\int_{s,y}\int_{K_1,K_2}e^{-i(P-A) \cdot s}e^{-iP\cdot (x_1-x_2)}e^{iK_1\cdot(x_1-y)+iK_2\cdot(y -x_2)}e^{iQ\cdot y},
\end{align}
with the last two factors from the Fourier transforms of two resummed correlation functions and EM fields respectively. Writing $iK_1\cdot x_1-iK_2\cdot x_2=\frac{i}{2}(K_1+K_2)\cdot(x_1-x_2)+\frac{i}{2}(K_1-K_2)\cdot (x_1+x_2)$, we find the $s$ and $y$ integrals fix $K_1$ and $K_2$ as
\begin{gather}
\frac{1}{2}(K_1+K_2)=P-A=K,\quad K_1=K_2+Q\nonumber\\
\Rightarrow\ K_1=K+\frac{Q}{2},\quad K_2=K-\frac{Q}{2}.
\end{gather}
The diagrammatic identity leads to the following representation
\begin{align}\label{O1Oq}
\frac{iK_1\cdot\bar{\s}}{K_1^2+i\h}(-i A\cdot\s)\frac{iK_2\cdot\bar{\s}}{K_2^2+i\h}.
\end{align}
We need to keep the $O(Q)$ contribution of \eqref{O1Oq}. The explicit representation of \eqref{O1Oq} indicates the effect of momentum inflow lies entirely in the shift of the momenta before and after the vertex. We expand the numerator and denominator of \eqref{O1Oq} respectively up to $O(Q)$
\begin{align}
\frac{1}{2}\tr[\s^\m K_1\cdot\bar{\s}A\cdot\s K_2\cdot\bar{\s}]&\simeq 2K^\m K\cdot A-A^\m K^2+i \e^{\m\n\r\s} K_\r Q_\s A_\n,\nonumber\\
\frac{1}{K_1^2+i\h}\frac{1}{K_2^2+i\h}&\simeq \frac{1}{(K^2+i\h)^2}+O(Q^2).
\end{align}
Note that the zeroth order $O(Q^0)$ contribution is already included in \eqref{O1}. We then easily obtain the remaining $O(Q)$ contribution as
\begin{align}\label{Oq}
O(Q):\;S_{ra}^\m=\frac{ \e^{\m\n\r\s} K_\r Q_\s A_\n}{(K^2+i\h)^2}.
\end{align}
The advanced counterparts of \eqref{O1} and \eqref{Oq} can be obtained from the above simply by flipping the sign of $i\h$.
To obtain the lesser correlation function, we need to use a generalized KMS relation in external EM fields. It is complicated by the momentum inflow from the EM fields. A detailed account of the KMS relation in external EM fields is given in Appendix~\ref{sec_app_B}. For the result at $O(Q^0)$, the KMS relation is formally the same as the free theory. 
\begin{align}\label{Sless_q0}
    O(Q^0):&\; S^{<\m}(K)=-(S^{ra}-S^{ar})\tilde{f}(k_0)=-2\p K^\m\d(K^2)\tilde{f}(k_0).
\end{align}
For the result at $O(Q)$, naive application of the prescription above encounters ambiguity in the choice of energy in the distribution function, which can be $k_0\pm\frac{1}{2}q_0$. However the difference only gives rise to a subleading contribution in the expansion in $Q$, thus to the order of our interest the KMS relation is formally the same. Using
\begin{align}
    \frac{1}{(K^2+i\h)^2}-\frac{1}{(K^2-i\h)^2}=2i\p\d'(K^2),
\end{align}
we obtain
\begin{align}\label{Sless_q1}
    O(Q):&\; S^{<\m}(K)=-2\p \e^{\m\r\n\s}K_\r Q_\s A_\n \d'(K^2)\tilde{f}(k_0).
\end{align}
The sum of \eqref{Sless_q0} and \eqref{Sless_q1} is in perfect agreement with the CKT result \eqref{Sless_CKT}.

Before  closing this section, we comment on two noteworthy aspects in the comparison:\\
1. The field theory results hold in equilibrium, while the CKT results hold in more general out-of-equilibrium case. The agreement is reached when specialized to the equilibrium case. In this case, the medium frame vector plays the role of a fictitious frame vector needed to define distribution function in CKT \cite{Chen:2014cla}.\\
2. The result \eqref{Sless_CKT} is based on collisionless CKT. Similarly, \eqref{Sless_q0} and \eqref{Sless_q1} involve no dissipation in the tree level calculations. These imply that the EM fields cannot do work on the particles, which would otherwise lead to instability\footnote{This restriction can be lifted when dissipation is included, see \cite{Gagnon:2006hi,Gagnon:2007qt} for dissipative transports in field theory approaches.}. The no-work condition $\vec{k}\cdot\vec{E}=0$ can be converted to covariant form by introducing a frame vector $n$ and defining the covariant EM fields as $E^\m=F^{\m\n}n_\n$ and $B^\m=\frac{1}{2}\e^{\m\n\r\s}n_\n F_{\r\s}$. The covariant no-work condition reads
\begin{align}
K\cdot E=0\to K\cdot Q\, \tilde{A}\cdot n-K\cdot \tilde{A}\, Q\cdot n=0,
\end{align}
where $\tilde{A}$ denotes the Fourier transform of the $A$ field. For arbitrary $A$ field, we need to have\footnote{The condition applies to fermion with specific momentum, which is appropriate for retarded correlation function involving no distribution function. For the lesser correlation function involving a distribution, this leads to the vanishing of electric field, which is the Killing condition in equilibrium in CKT.}
\begin{align}\label{Q_cons}
K\cdot Q=0,\quad Q\cdot n=0.
\end{align}
The second condition of \eqref{Q_cons} implies that the resulting electric field is static, and the first condition indicates that the electric field is perpendicular to particle momentum.


\section{In-medium electromagnetic form factors}\label{sec_3}

We are ready to reformulate the results in the previous section using EM form factors in vacuum. We start with the representation \eqref{O1Oq} reproduced for convenience below
\begin{align}\label{rep}
S_{ra}=\frac{iK_1\cdot\bar{\s}}{K_1^2+i\h}(-i A\cdot\s)\frac{iK_2\cdot\bar{\s}}{K_2^2+i\h},
\end{align}
where $K_1=K+\frac{Q}{2}$ and $K_2=K-\frac{Q}{2}$ are kinetic momenta and $A$ corresponding to the insertion of background gauge field with momentum inflow. Now we will show that \eqref{rep} can be interpreted as scattering of particle off EM fields if we combine the on-shell condition $K^2=0$ and the kinematic condition \eqref{Q_cons}.

With $K^2=0$ and $K\cdot Q=0$ from \eqref{Q_cons} (we will use the other condition of \eqref{Q_cons} shortly), we easily find $K_{1,2}$ are on-shell up to $O(Q^2)$: $K_{1,2}^2=K^2\pm K\cdot Q+\frac{Q^2}{4}\sim O(Q^2)$, thus we can interpret $K_2$ and $K_1$ as incoming and outgoing momenta respectively. To see that it corresponds to scattering off EM fields, we use the spinor representation for right-handed fermions with positive energy
\begin{align}
K_i\cdot\bar{\s}=u(K_i)u^\dg(K_i),
\end{align}
with $i=1,2$ to rewrite \eqref{rep} as
\begin{align}\label{scat}
\frac{iu(K_1)}{K_1^2+i\h}\big[u^\dg(K_1) i\s^\m u(K_2)A_\m\big]\frac{iu^\dg(K_2)}{K_2^2+i\h}.
\end{align}
The expression in the square bracket is readily identified as an amplitude for scattering of right-handed particle off EM fields. It follows that the correlation function adopts more general description in terms of EM form factors.

The choice of EM form factors is not unique, we will define EM form factors that are convenient for physics of spin polarization. Taking $n^\m=(1,0,0,0)$ and using the following explicit relations \cite{Dong:2021fxn}
\begin{align}\label{spinor_products}
&u^\dg(K_1)u(K_2) = \x_1^\dg\x_2(4k_1k_2)^{1/2},\nonumber\\
&u^\dg(K_1)\s^i u(K_2) = \x_1^\dg\x_2(4k_1k_2)^{1/2}\frac{k_1k^i_2+k^i_1k_2+i\e^{ijk}k_{2j}k_{1k}}{k_1k_2+\vec{k}_1\cdot\vec{k}_2},
\end{align}
with $k_{1,2}=|\vec{k}_{1,2}|$, we can rewrite the square bracket in \eqref{scat} into a more instructive form up to $O(Q)$
\begin{align}
u^\dg(K_1) \(n^\m+\hat{k}^\m+\frac{i\e^{\m\n\r\s} n_\n K_\r Q_\s}{2(K\cdot n)^2}\) u(K_2),
\end{align}
with $\hat{k}=(0,\frac{\vec{k}}{|\vec{k}|})$ and $\vec{k}=\vec{K}_1+\vec{K}_2$. The three structures are readily identified as temporal, longitudinal and transverse parts of the vertex. One may check that they satisfy the Ward identity as $Q\cdot n=0$, $Q\cdot\hat{k}=\frac{Q\cdot K}{K\cdot n}-Q\cdot n=0$ and $Q_\m \e^{\m\n\r\s} n_\n K_\r Q_\s=0$ by using the kinematic conditions \eqref{Q_cons}. We can formally write the vertex function as\footnote{The reason we are able to parameterize the vertex using structures without Dirac structure is that the number of degree of freedom for right-handed particle is one with spin enslaved by momentum.}
\begin{align}
\G^\m=F_0u^\m+F_1\hat{k}^\m+F_2\frac{i\e^{\m\n\r\s}n_\n K_\r Q_\s}{2(K\cdot n)^2},
\end{align}
with $F_0=F_1=F_2=1$, that is at tree level, all form factors are degenerate\footnote{In fact the degeneracy remains true in the presence of quantum fluctuation in vacuum for Weyl spinors, cf \cite{Giunti:2014ixa}.}.

With these preparations, the extension to in-medium EM form factors is minimal. Since the medium introduces a preferred frame, it is convenient to use the medium frame vector $u^\m$ to replace the arbitrary frame vector $n^\m$. Thus the vertex can be parameterized by three EM form factors as
\begin{align}
\G^\m=F_0u^\m+F_1\hat{k}^\m+F_2\frac{i\e^{\m\n\r\s}u_\n K_\r Q_\s}{2(K\cdot u)^2} \label{parameterization}
\end{align}
with $F_0$, $F_1$ and $F_2$ being the dimensionless EM form factors.

The physical meaning of the EM form factors can be seen by calculating the correlation function. With $\G^\m$ being the effective vertex in the presence of medium, we expect the corresponding correlation function to be given by the replacement $\s^\m\to\G^\m$ in \eqref{rep} (we show the validity of replacement show in Appendix~\ref{sec_app_C})
\begin{align}
S_{ra}=\frac{iK_1\cdot\bar{\s}}{K_1^2+i\h}(-i A\cdot\G)\frac{iK_2\cdot\bar{\s}}{K_2^2+i\h}.
\end{align}
The trace can be evaluated as
\begin{align}\label{trace}
&\frac{1}{2}\tr[K_1\cdot\bar{\s}A\cdot\G K_2\cdot\bar{\s}]\simeq F_2(\vec{B}\cdot\vec{k}),\\ &\frac{1}{2}\tr[\s^i K_1\cdot\bar{\s}A\cdot\G K _2\cdot\bar{\s}]\simeq \(F_0\e^{ijk}E_jk_k+F_1(k_0B^i-\vec{B}\cdot \vec{k}\hat{k}^i)+F_2(\vec{B}\cdot\vec{k})\hat{k}^i\).
\end{align}
We have kept the $O(Q)$ terms only with the same reasoning as in the previous section and used the condition $Q\cdot u=0$ from \eqref{Q_cons}. We show in Appendix~\ref{sec_app_C} that the KMS relation remain valid at $O(Q)$ in the presence of medium interaction. Using the KMS relation, we arrive at finally
\begin{align}
S^{<0 } &= F_2 \left(\vec{k} \cdot \vec{B}\right) 2\pi \delta '(K^2) \tilde{f}(k_0) ,\label{0_physical_meanings_of_form_factors}\\
S^{<i } &= \left[F_0 \epsilon^{ijk}E_j k_k + F_1 \left( k_0 B^i - \left(\vec{B} \cdot \vec{k}\right)\hat{k}^i\right) + F_2 \left(\vec{B} \cdot \vec{k}\right)\hat{k}^i\right] 2\pi \delta '(K^2) \tilde{f}(k_0). \label{i_physical_meanings_of_form_factors}
\end{align}
In (\ref{0_physical_meanings_of_form_factors}) and (\ref{i_physical_meanings_of_form_factors}), we see that the three form factors give rise to different spin couplings: $F_0$ corresponds to the spin Hall effect; $F_1$ corresponds to the spin-perpendicular magnetic coupling, and $F_2$ corresponds to the spin-parallel magnetic coupling. We expect interaction with medium to lift the degeneracy of the form factors in vacuum. We will present an example in the next section.
%
%

We close this section by discussing two general properties of the EM form factors:\\
1. We have focused on right-handed particle in defining the EM form factors. For left-handed particle, the $F_2$ term should flip sign. 
Note that in this case, components of Wigner function is given by (instead of \eqref{Sless_CKT})
\begin{align}
S^{<\m}(X,K)=\frac{1}{2}\tr[\bar{\s}^\m S^<(X,K)].
\end{align}
It is not difficult to find the counterparts of \eqref{0_physical_meanings_of_form_factors} and \eqref{i_physical_meanings_of_form_factors} should also be multiplied by $-1$ for left-handed particle. It follows that in the absence of chiral imbalance, $\tilde{f}_R(k_0)=\tilde{f}_L(k_0)$, the axial component of Wigner function from the difference of right-handed and left-handed contributions is twice those of \eqref{0_physical_meanings_of_form_factors} and \eqref{i_physical_meanings_of_form_factors}.
\\
2. It is instructive to analyze the property of the EM form factors under time-reversal symmetry. Note that $\G^i$ is roughly speaking the spin (as $\G^i$ replaces $s\s^i$), which is T-odd. It follows that the EM form factors are all T-even. Since in medium frame $q_0=0$, the EM form factors can only a function of $k^2$ and $q^2$. Thus we expect the EM form factors to be real functions. We shall confirm this by an explicit example \footnote{In the presence of dissipation, the EM form factors also depend on relaxation time, which is T-odd, the corresponding EM form factors are complex in general.}.

\section{QCD radiative corrections to EM form factors}\label{sec_4}

As an example, we consider modification to the vacuum EM form factors in a QGP medium from QCD radiative corrections. Similar to the vacuum case, loop diagrams lead to vertex corrections as well as self-energy corrections, corresponding to modifications to interaction and incoming/outgoing states respectively. We shall absorb both corrections into in-medium EM form factors. Since the aim of this section is a medium modified vertex, we can restrict ourselves to $O(A)$. It follows that there is no difference between canonical and kinetic momenta. We shall use $P_{1,2}$ to denote incoming and outgoing momenta.
%

\begin{figure}
  \centering
    \includegraphics[width=.8\textwidth]{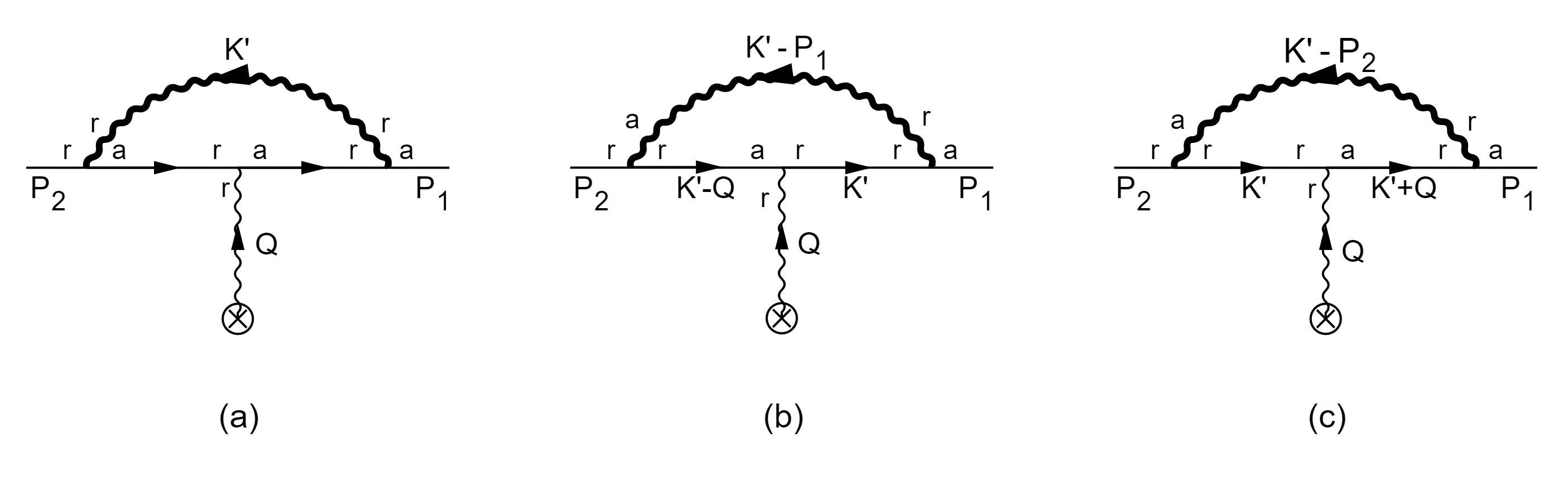}
  \caption{One-loop diagrams for vertex corrections with three possible labelings in $ra$ (gluons are represented by thick curves)}
  \label{vertex_correction}
\end{figure}

We proceed in the $ra$-basis. The one-loop vertex corrections come from three diagrams shown in Figure~\ref{vertex_correction}.
We need the following propagators in $ra$-basis in addition to \eqref{S_prop1}-\eqref{S_prop3}
\begin{align}
   S_{rr}(P)&=-2\pi \slashed{P}\varepsilon(p_0)\left(\tilde{f}(p_0)-\frac{1}{2}\right)\delta (P^2), \\
    S_{ra}(P)&=\frac{i\slashed{P}}{P^2+i \eta\,\varepsilon(p_0)},\\
    S_{ar}(P)&=\frac{i\slashed{P}}{P^2-i \eta\,\varepsilon(p_0)}, \\
   D^{\mu \nu }_{rr}(P)&=-2\pi g^{\mu \nu }\varepsilon(p_0)\left(f(p_0)+\frac{1}{2}\right)\delta (P^2), \\
   D^{\mu \nu }_{ra}(P)&=\frac{-ig^{\mu \nu }}{P^2+i \eta \varepsilon(p_0)}, \\
   D^{\mu \nu }_{ar}(P)&=\frac{-ig^{\mu \nu }}{P^2-i \eta \varepsilon(p_0)},
\end{align}
where $\slashed{P}=P\cdot\gamma$, $f(k_0)$ is the Bose-Einstein distribution function and $D^{\mu\nu}$ denotes the boson propagator. We have chosen the Feynman gauge $(\xi=1)$ for the gluon propagators and suppressed the color factors. 
To further simplify the calculation, we keep only the leading medium contributions to the EM form factors by using the hard thermal loop (HTL) approximation. The factor $\frac{1}{2}$ in $S_{rr}$ and $D^{\mu \nu }_{rr}$ will be ignored as the vacuum contribution. Diagram (a) of Figure \ref{vertex_correction} is evaluated as:
\begin{align}
   &\int \frac{d^{4} K'}{( 2\pi )^{4}}\left( -ig\gamma ^{\rho } T_{ij}^{a}\right)\frac{i(\slashed{K}'+\slashed{P}_1)}{(K'+P_1)^{2}-i \varepsilon(k'_0+p_1^0)\eta }\left( -i\gamma ^{\nu } \right)\frac{i(\slashed{K}'+\slashed{P}_2)}{(K'+P_2)^{2}-i \varepsilon(k'_0+p_2^0)\eta }\left( -ig\gamma ^{\sigma } T_{ij}^{a}\right)\nonumber\\
   &\times( -g_{\rho \sigma }) 2\pi \delta \left( K'^{2}\right) f( k'_0)A_{\nu },
\end{align}
where $\left( -ig\gamma ^{\rho } T_{ij}^{a}\right)$ denotes the fermion-gluon vertex. In the HTL approximation, we ignore $P_2$ and $P_1$ next to $K'$ in the numerators. The denominators can be simplified using $\d(K'^2)$ and on-shell conditions of $P_2$ and $P_1$. We then have
\begin{align}\label{diag_a}
   -2ig^2C_F A_\nu \int\frac{d^{4} K'}{( 2\pi )^{4}} 2K'^\nu \slashed{K}' \frac{1}{2K'\cdot P_1-i \varepsilon(k'_0)\eta}\frac{1}{2K'\cdot P_2-i \varepsilon(k'_0)\eta}2\pi \delta (K'^2)f(k'_0).
\end{align}
There is potential divergence when the momenta $K'$ and $P_1$($P_2$) become collinear. The collinear divergence does not contribute to EM form factors as we shall see below, allowing us to ignore $i\h$ here. Similarly diagrams (b) and (c) are evaluated as
\begin{align}\label{diag_b}
   (b)&:\int \frac{d^{4} K'}{( 2\pi )^{4}}\left( -ig\gamma ^{\rho } T_{ij}^{a}\right)\slashed{K}'\left( -i\gamma ^{\nu } A_{\nu }\right)\frac{i(\slashed{K}'-\slashed{Q})}{(K'-Q)^{2}+i \eta }\left( -ig\gamma ^{\sigma } T_{ij}^{a}\right)\frac{-ig_{\rho \sigma }}{(K'+P_1)^{2}-i \eta } 2\pi \delta \left( K'^{2}\right) \tilde{f}( K'_{0}) \nonumber\\
   &\simeq 2ig^2C_F A_\nu \int\frac{d^{4} K'}{( 2\pi )^{4}} 2K'^\nu \slashed{K}' \frac{1}{2K'\cdot Q}\frac{1}{2K'\cdot P_1}2\pi \delta (K'^2)\tilde{f}(k'_0),
\end{align}
and
\begin{align}\label{diag_c}
   (c)&:\int \frac{d^{4} K'}{( 2\pi )^{4}}\left( -ig\gamma ^{\rho } T_{ij}^{a}\right)\frac{i(\slashed{K}'+\slashed{Q})}{(K'+Q)^{2}-i \eta }\left( -i\gamma ^{\nu } A_{\nu }\right) \slashed{K}' \left( -ig\gamma ^{\sigma } T_{ij}^{a}\right)\frac{-ig_{\rho \sigma }}{(K'-P_2)^{2}-i \eta } 2\pi \delta \left( K'^{2}\right) \tilde{f}( K'_{0}) \nonumber\\
   &\simeq 2ig^2C_F A_\nu \int\frac{d^{4} K'}{( 2\pi )^{4}} 2K'^\nu \slashed{K}' \frac{1}{2K'\cdot Q}\frac{1}{2K'\cdot P_2}2\pi \delta (K'^2)\tilde{f}(k'_0).
\end{align}
Here $Q^2$ are dropped in the denominators as $O(Q^2)$ subleading corrections.
Summing over \eqref{diag_a}, \eqref{diag_b} and \eqref{diag_c}, we obtain
\begin{align}\label{diag_abc}
   &-2ig^2C_F A_\nu \int\frac{d^{4} K'}{( 2\pi )^{4}} 2K'^\nu \slashed{K}' \left[ \frac{1}{2K'\cdot P_1}\frac{1}{2K'\cdot P_2}\(f(k'_0)+\tilde{f}(k'_0)\) \right]2\pi \delta (K'^2) \nonumber\\
   &=-2ig^2C_F A_\nu \int\frac{d^{4} K'}{( 2\pi )^{4}} 2K'^\nu \slashed{K}'\left(\frac{1}{2K'\cdot P_2}-\frac{1}{2K'\cdot P_1}\right)\frac{1}{2K'\cdot Q}\left(f(k'_0)+\tilde{f}(k'_0)\right)2\pi \delta (K'^2). 
\end{align}
We note that the representation similar to first line of \eqref{diag_abc} exists in imaginary time in \cite{le2000thermal}, but analytic continuation to real time is quite non-trivial for vertex function and there is hidden IR divergence, which we treat carefully below.

We first look at the collinear divergences using the second line of \eqref{diag_abc}. The two terms in the second bracket of \eqref{diag_abc} are simply related by $P_1\leftrightarrow P_2$. In fact the collinear divergence does not contribute to EM form factors for the following reason: since the vertex function is sandwiched between incoming/outgoing 
fermion states as $\bar{u}(P_1)\G^\m u(P_2)$, terms like $\slashed{P}_2$ and $\slashed{P}_1$ simply does not contribute to the amplitude by equation of motion (EOM). So we may subtract appropriate structures to eliminate the collinear divergence as
\begin{align}
&-2ig^2C_F A_\nu \int\frac{d^{4} K'}{( 2\pi )^{4}} 2 K'^\nu \left(\slashed{K}' -\frac{K' \cdot u}{P_2 \cdot u}\slashed{P}_2 \right) 
\frac{1}{2K'\cdot P_2}\frac{1}{2K'\cdot Q}\left(f(k'_0)+\tilde{f}(k'_0)\right)2\pi \delta (K'^2)\nonumber\\
&-(P_2\to P_1) ,\label{integral_with_counterterm}
\end{align}
where $u^\mu =(1, 0, 0, 0)$ denotes the QGP frame vector. The subtracted structure ensures that $K'^\lambda -\frac{K' \cdot u}{P_{i} \cdot u}P_{i}^\lambda$ and $2K'\cdot P_{i}$ approach zero at the same rate, making the integrand regular in the collinear limit. There is actually another divergence occurring when $Q\to0$ (note that $q_0=0$ from \eqref{Q_cons}). This is an IR divergence, which will be cut off by screening effect of the medium.

To perform the tensor integrals, we adopt the following parameterization of the momenta by using the rotational symmetry:
\begin{align}
   P_1 &=(p_0, 0, \frac{q}{2}, p) ,\nonumber \\
   P_2 &=(p_0, 0, -\frac{q}{2}, p) ,\nonumber \\
   Q &=(0, 0, q, 0) ,\nonumber \\
   K' &=(k'_0, k' \sin{\theta } \cos{\varphi }, k' \sin{\theta } \sin{\varphi }, k' \cos{\theta }).\label{parameterization}
\end{align}
where $p$, $q$ denote the lengths of $\vec{p}$, $\vec{q}$. One of the advantages of this parameterization is that many of the components in the integral can be eliminated, leaving only those which have definite physical meaning.
After some algebra, we can see that most of the components in (\ref{integral_with_counterterm}) are zero after integration\footnote{The components with $\lambda = 2$ can be written as $Q\cdot \sigma$ in the parameterization (\ref{parameterization}), which also have no contribution to the EM form factors by the EOM.  }, leaving only $(\lambda, \nu)=(1, 1), (3, 3), (3, 0)$: 
\begin{align}
(\ref{integral_with_counterterm}) = \left\{
\begin{aligned}
-2i A_\nu \frac{m_f^2}{4p^2}\ln{\frac{2p}{q}}, \quad\quad\quad\quad   & (\lambda,\nu)=(1,1)  \\
-2i A_\nu \frac{m_f^2}{4p^2}\left(1-\ln{\frac{2p}{q}}\right), \quad\quad   & (\lambda,\nu)=(3,3)  \\
-2i A_\nu \frac{m_f^2}{4p^2}\left(-\ln{\frac{2p}{q}}\right), \quad\quad   & (\lambda,\nu)=(3,0)
\end{aligned}
\right.
\end{align}
with
\begin{align}
m_f^2=g^2C_F\int\frac{4\p kdk}{(2\pi)^3}\left(f(k_0)+\tilde{f}(k_0)\right)=\frac{1}{8}g^2 T^2 C_F,
\end{align}
is the thermal mass of fermions with the group factor $C_F=\frac{4}{3}$.
Defining $\hat{l}^i=\frac{1}{pq}\epsilon^{ijk}q_j p_k$, we can rewrite the integral in (\ref{integral_with_counterterm}) in a covariant form:
\begin{align}
-2i m_f^2 A_\nu \gamma _\lambda \left( \hat{l}^\lambda \hat{l}^\nu \frac{1}{p^2}\ln{\frac{2p}{q}} + \hat{p}^\lambda \hat{p}^\nu \frac{1}{p^2} - \hat{p}^\lambda P^\nu \frac{1}{p^3}\ln{\frac{2p}{q}} \right), \label{vertex_without_thermal_mass}
\end{align}
with $\hat{p}^\nu = (0,\frac{\vec{p}}{p})$.
 
As mentioned earlier, (\ref{vertex_without_thermal_mass}) still contains IR divergences as $q\to0$. To see the origin of the divergence, it is instructive to rewrite part of the integrand of \eqref{diag_abc} as\footnote{We could have used this representation directly in \eqref{diag_abc}, but this would make the origin of the divergence less clear.}
\begin{align}
\(\frac{1}{2K'\cdot P_2}-\frac{1}{2K'\cdot P_1}\)\frac{1}{2K'\cdot Q}=\frac{1}{2K'\cdot P_2}\frac{1}{2K'\cdot P_1}.
\end{align}
We see the cancellation of the factors $2K'\cdot Q$ in the numerator and denominator. This implies partial cancellation of the IR divergence among diagram (b) and (c). The remaining divergence arise in the kinematic region where $P_1$ and $P_2$ become collinear with $K'$ simultaneously. To divergence can be cut off by screening effect of the medium. The screening effect can be implemented by using dressed propagators. The discussion above suggest that we only need to dress the propagators with momenta $K'\pm P_1$ and $K'\pm P_2$. This amounts to the following substitutions \cite{le2000thermal}
\begin{align}
\frac{1}{2K'\cdot P_{1,2}} \to \frac{1}{2K'\cdot P_{1,2}+2m_f^2} \label{substitute_1}
\end{align}
for fermions, and
\begin{align}
\frac{-g_{\mu \nu }}{2K'\cdot P_{1,2}} \to \frac{\mathcal{P}^T_{\mu \nu }}{2K'\cdot P_{1,2}+m_g^2} \label{substitute_2}
\end{align}
for gluons.\footnote{The longitudinal polarization does not contribute in the HTL approximation, leaving only the transverse polarization with $\mathcal{P}^T_{\mu \nu }$ being the transverse projector. The change of polarization tensor does not change the Dirac structure \cite{Lin:2023ass}.} Here $m_g^2=\frac{1}{3}g^2T^2(C_A+\frac{1}{2}N_f)$ is the gluon thermal mass, with $C_A=3$ and $N_f$ being the number of flavors in QGP.
 After applying (\ref{substitute_1}) and (\ref{substitute_2}) to the internal propagators in (\ref{diag_abc}), we obtain:
\begin{align}
\begin{split}
-2ig^2C_F A_\nu \int\frac{d^{4} K'}{( 2\pi )^{4}} 2K'^\nu \slashed{K}' \Big[ \frac{1}{2K'\cdot P_1+2m_f^2}\frac{1}{2K'\cdot P_2+2m_f^2}f(k'_0) \\
+ \left(\frac{1}{2K'\cdot P_2+m_g^2}-\frac{1}{2K'\cdot P_1+m_g^2}\right)\frac{1}{2K'\cdot Q}\tilde{f}(k'_0) \Big] 2\pi \delta (K'^2) ,
\end{split}
\end{align}
By repeating the previous procedure for integration, we can finally arrive at the vertex corrections to the vertex function $\delta \Gamma^\nu_{vertex}$:
\begin{align}\label{vertex_corr}
\delta \Gamma_{vertex} ^\nu A_\nu \Rightarrow&\ 2 m_f^2 A_\nu \gamma _\lambda \Bigg{[} \frac{1}{6 p^2}\(2 \ln \left(\frac{pT}{m_f^2}\right)+\ln \left(\frac{2 pT}{m_{g }^2}\right) 
-36 \ln (A)+\ln \left(16 \pi ^3 \right)+3\)\nonumber\\
&\times\left(\hat{l}^\lambda \hat{l}^\nu - \hat{p}^\lambda P^\nu \frac{1}{p} \right) + \hat{p}^\lambda \hat{p}^\nu \frac{1}{p^2} \Bigg{]},
\end{align}
with $A\simeq 1.282$ being the Glaisher constant. We can see the thermal masses of fermions and gluons effectively cut off the logarithmic divergences.

\begin{figure}\label{self-energy}
  \centering
    \includegraphics[width=.8\textwidth]{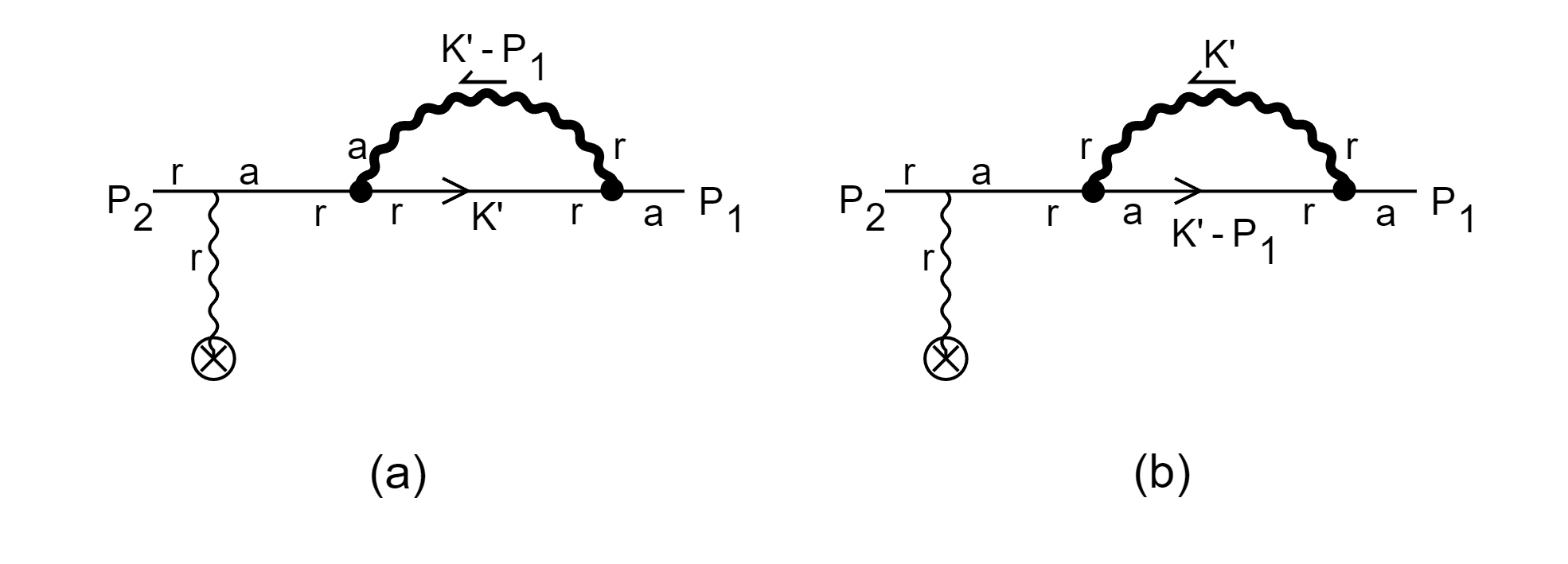}    
    \includegraphics[width=.8\textwidth]{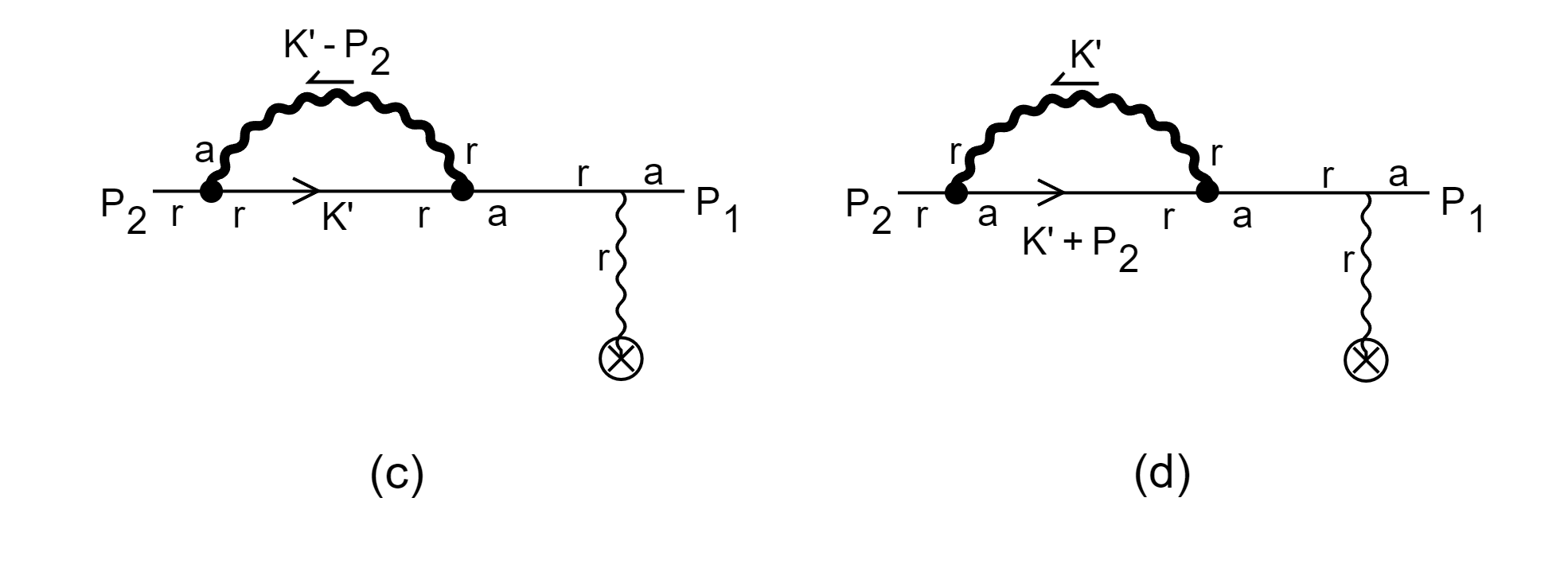}
  \caption{self-energy corrections to both incoming and outgoing states, each with two different labelings in $ra$.}
\end{figure}

The remaining contribution to the EM form factors is from the self-energy corrections to the incoming/outgoing states. The corresponding diagrams are collected in Figure~\ref{self-energy}.
These self-energy corrections are actually similar to the vacuum counterpart. That is, the self-energy corrections can be absorbed into the field strength renormalization. Under the condition $m_f^2 \ll p \ll T$, we have the self-energy corrections to the vertex function $\delta \Gamma^\nu_{self-energy}$ \cite{le2000thermal}:
\begin{align}\label{self_corr}
\delta \Gamma_{self-energy} ^\nu A_\nu \Rightarrow 2 \frac{m_f^2}{2p^2}\left( 1-ln\frac{2p^2}{m_f^2} \right) {\slashed A}.
\end{align}

Now we can combine \eqref{vertex_corr} and \eqref{self_corr}, and sandwich the resulting $\d\G^\m$ between $u^\dg(K_1)$ and $u(K_2)$ to find the leading medium corrections to the three form factors:
\begin{align}
\delta F_0 &= \frac{2m_f^2}{k^2}X+\frac{m_f^2}{k^2}\left( 1-ln\frac{2k^2}{m_f^2} \right), \label{delta_F_0}\\
\delta F_1 &= \frac{2m_f^2}{k^2}(X-1)+\frac{m_f^2}{k^2}\left( 1-ln\frac{2k^2}{m_f^2} \right), \label{delta_F_1}\\
\delta F_2 &= \frac{2m_f^2}{k^2}X+\frac{m_f^2}{k^2}\left( 1-ln\frac{2k^2}{m_f^2} \right), \label{delta_F_2}
\end{align}
where $$X=\frac{1}{6}\left(2 \ln \left(\frac{kT}{m_f^2}\right)+\ln \left(\frac{2 kT}{m_{g}^2}\right)-36 \ln (A)+\ln \left(16 \pi ^3 \right)+3\right),$$
and we have replaced $p$ with the kinetic momentum $k$ to account for effect of the external EM fields at $\mathcal{O}(Q^0)$.
Let us discuss the spin polarization effect based on \eqref{i_physical_meanings_of_form_factors}:\\
i. In vacuum all three form factors are degenerate as a consequence of Lorentz symmetry. In the medium, the degeneracy is expected to be lifted. Interestingly, our explicit results show partial lift with $\d F_1\ne\d F_2=\d F_0$, indicating the spin couplings to parallel and perpendicular magnetic fields are different, but the spin couplings to perpendicular electric and parallel magnetic fields remain the same. We believe the remaining degeneracy is a consequence of the HTL approximation, which may be fully lifted beyond the approximation.\\
ii. The medium corrections to all EM form factors are real, which is consistent with our analysis based on time-reversal symmetry. This can be traced back to our requirement that the EM fields do no work on the fermions, so that no dissipation is involved. Relaxing the requirement would necessarily lead to complex EM form factors, but also with more complicated structures of the vertex.\\
iii. We have not considered self-energy corrections to EOM. It is actually not needed: because chiral symmetry is unbroken and respected by the self-energy $\S\propto \slashed{P}$ \footnote{It is true for the particle excitation we consider, but not so for plasmino \cite{le2000thermal}} thus the self-energy corrected EOM is the same as the free EOM up to an overall normalization. This justifies the use of spinor solution to the free EOM as the incoming/outgoing states.\\
%
iv. The results \eqref{delta_F_1} and \eqref{delta_F_2} show that $F_2>F_1$, i.e. the spin coupling to parallel magnetic field is larger than the coupling to perpendicular magnetic field. This has an interesting implication for spin polarization in HIC. Strictly speaking the results obtained with massless fermions cannot be directly applied to spin polarization measured in rest frame of particles in experiments, see however an extrapolation to massive fermions in \cite{Yi:2021ryh}. We will use $S_R^{<i}-S_L^{<i}$ as a proxy for ``spin polarization in the medium frame'' for illustration purpose. In the absence of medium interaction, we have $F_0=F_1=F_2$. As we have discussed at the end of the previous section
\begin{align}\label{splitting}
S_R^{<i}-S_L^{<i}=2S_R^{<i}\propto B^i
\end{align}
from the SBC for particle in the absence of chiral imbalance. Because the spin magnetic couplings for particle and anti-particle differ by the sign, \eqref{splitting} also corresponds to the splitting of polarization between particle and anti-particles. This has the same origin as the splitting between $\L$ and anti-$\L$ hyperon global polarization \cite{Becattini:2016gvu}. The global nature lies in the fact that the magnetic field induced polarization is always along the field direction. Now with medium interaction lifting the degeneracy of $F_1$ and $F_2$, \eqref{splitting} is modified to
\begin{align}\label{splitting2}
S_R^{<i}-S_L^{<i}=2S_R^{<i}\propto F_1k_0B^i+(F_2-F_1)(\vec{B}\cdot\vec{k})\hat{k}^i.
\end{align}
We see that the magnetic field induced polarization contains a momentum dependent component. It is tempting to interpret the term proportional to $F_2-F_1$ as local polarization. Such a component might be present in the splitting between $\L$ and anti-$\L$ hyperon polarization. Further phenomenological studies are needed to confirm this effect.



\section{Summary and outlook}\label{sec_outlook}

We have rederived the fermion correlation function in external electromagnetic fields obtained in chiral kinetic theory using field theory by resumming tree level diagrams. Along the way, we identify kinematic conditions that ensure the electric field does no work to fermions. The kinematic conditions allow us to interpret the correlation function as scatterings amplitude of particle off electromagnetic fields. We define three electromagnetic form factors that correspond to spin coupling to electric field, perpendicular and parallel magnetic fields respectively. We calculate QCD radiative corrections to the form factors in QGP under the HTL approximation as an example. We find the radiative corrections lead to partial lift of the degeneracy of the form factors at tree level. In particular, the spin couples to parallel and perpendicular magnetic fields differently, while the spin couples to perpendicular electric and magnetic field with the same magnitude. The results indicate that the magnetic field induced polarization also contains a momentum dependent component. With a caveat in mind, we suggest there might be local polarization in the splitting between $\L$ and anti-$\L$ hyperons in heavy ion collisions \cite{Becattini:2016gvu}. Further phenomenological studies are needed to confirm this effect.

On theoretical perspective, it would be interesting to relax the kinematic restrictions to study spin coupling with more general electromagnetic fields. This will necessarily lead to more structures of interaction vertex and complex form factors, which will enable us to study spin polarization as well as spin relaxation phenomenon.

While the bulk of the analysis in this paper relies on perturbation theory, the interpretation of scattering and formulation using form factors is not restricted to weakly interacting systems and also not restricted to elementary particles. An interesting application is on composite particles. It is known that the vacuum electromagnetic form factors characterize the charge and spin distributions of the composite particle as seen by external electromagnetic fields. In-medium electromagnetic form factors encodes how the medium modifies the charge and spin distributions. We leave this for future studies.

\section*{Acknowledgments}
This work is in part supported by NSFC under Grant Nos 12075328, 11735007.

\appendix

\section{KMS relation in external electromagnetic fields}\label{sec_app_B}

In this appendix, we derive the KMS relation in the presence of slow-varying EM fields. We shall see up to $O(q)$ it is formally the same the usual KMS relation. From \eqref{Wigner}, we know the resummed correlation function and the counterpart without EM fields are related by
\begin{align}
S^{ra}(K)=S_{re}^{ra}(P=K+A),
\end{align}
for constant $A$ field. The same relation actually holds for correlation functions with other operator orderings. To see this, we note that the effect of constant $A$ on correlation function can be easily deduced from the Lagrangian ${\cal L}=\bar{\psi}(i\pd_\m-A_\m)\s^\m\psi$. With $i\pd_\m$ identified as canonical momentum $P_\m$, we see constant $A$ simply changes canonical momentum into kinetic one $K=P-A$. Furthermore, the gauge link in the correlation function fixes $K$ as argument of Wigner function. Therefore the usual KMS relation in free theory holds for the resummed correlation function. In particular
\begin{align}
S_{re}^<(P)=-e^{-\b k_0}S_{re}^>(P)\Rightarrow S^<(K)=-e^{-\b k_0}S^>(K).
\end{align}

At $O(Q)$, as discussed in Sec~\ref{sec_2}, we shall keep the $O(Q)$ contribution of the expression below
\begin{align}
S^<_{re}(K) &= -\int d^4K_1 \left[ S_0^<(K_1)(-iA\cdot \sigma )S_0^F(K_2)-S_0^{\bar{F}}(K_1)(-iA\cdot \sigma )S_0^<(K_2) \right] \delta ^4(K_2+Q-K_1) \nonumber\\
&= -\left[S_0^<(K_1)(-iA\cdot \sigma )S_0^F(K_2)-S_0^{\bar{F}}(K_1)(-iA\cdot \sigma )S_0^<(K_2)\right]_{K_1\to K_2+Q},
\end{align}
with the following resummed correlation function at $O(Q^0)$
\begin{equation}
\begin{aligned}
S^F_0 &= K\cdot \bar{\sigma}\left[ \frac{i}{K^2+i \eta }-2\pi \tilde{f}(k_0)\delta (K^2) \right] ,\\
S^{\bar{F}}_0 &= K\cdot \bar{\sigma}\left[ \frac{-i}{K^2-i \eta }-2\pi \tilde{f}(k_0)\delta (K^2) \right] ,\\
S^<_0 &= -2\pi K\cdot \bar{\sigma}\tilde{f}(k_0)\delta (K^2) ,\\
S^>_0 &= -2\pi K\cdot \bar{\sigma}(1-\tilde{f}(k_0))\delta (K^2).
\end{aligned}
\end{equation}
We first find a vanishing contribution from the second terms of $S^F$ and $S^{\bar{F}}$:
\begin{align}
\left[ 2\pi \tilde{f}(k_1^0)\delta (K_1^2)2\pi \tilde{f}(k_2^0)\delta (K_2^2)-2\pi \tilde{f}(k_1^0)\delta (K_1^2)2\pi \tilde{f}(k_2^0)\delta (K_2^2) \right](K_1 \cdot \bar{\sigma} )(-iA \cdot \sigma )(K_2 \cdot \bar{\sigma} ) = 0,
\end{align}
which means that in $S^F$ and $S^{\bar{F}}$ only the vacuum part contribute to the resummed correlation function, thus we have 
\begin{align}\label{resum_KMS}
&S^<_{re}(K) = -\left[S_0^<(K_1)(-iA\cdot \sigma )\frac{iK_2\cdot \bar{\sigma }}{K_2^2+i \eta }-\frac{-iK_1\cdot \bar{\sigma }}{K_1^2-i \eta }(-iA\cdot \sigma )S_0^<(K_2)\right]_{K_1\to K_2+Q}\nonumber\\
&=-2\p\left[-\tilde{f}(k_1^0)\d(K_1^2)K_1\cdot\bar{\s}(-iA\cdot \sigma )\frac{iK_2\cdot \bar{\sigma }}{K_2^2+i \eta }-\frac{-iK_1\cdot \bar{\sigma }}{K_1^2-i \eta }(-iA\cdot \sigma )K_2\cdot\bar{\s}\tilde{f}(k_2^0)\d(K_2^2)\right]_{K_1\to K_2+Q},
\end{align}
where we have used the KMS relation for free propagators in the second line.
We shall relate the above to $S_{re}^{ra}$ and $S_{re}^{ar}$, which are given by
\begin{align}
&S_{re}^{ra}=\frac{iK_1\cdot\bar{\s}}{K_1^2+i\h}(-iA\cdot\s)\frac{iK_2\cdot\bar{\s}}{K_2^2+i\h},\nonumber\\
&S_{re}^{ar}=\frac{iK_1\cdot\bar{\s}}{K_1^2-i\h}(-iA\cdot\s)\frac{iK_2\cdot\bar{\s}}{K_2^2-i\h}.
\end{align}
Using the following identity
\begin{align}
&\frac{1}{K_1^2+i\h}\frac{1}{K_2^2+i\h}-\frac{1}{K_1^2-i\h}\frac{1}{K_2^2-i\h}\nonumber\\
=&\(-2i\p\d(K_1^2)\)\frac{1}{K_2^2+i\h}-\frac{1}{K_1^2-i\h}\(2i\p\d(K_2^2)\),
\end{align}
we easily find the KMS relation holds if $\tilde{f}(k_1^0)=\tilde{f}(k_2^0)$. Recall that we are supposed to keep the $O(Q)$ contributions, thus we may replace $k_1^0$ and $k_2^0$ by $k_0$ to obtain
\begin{align}
S_{re}^<(K)=-\tilde{f}(k_0)(S_{re}^{ra}(K)-S_{re}^{ar}(K)),
\end{align}
with a suppressed $O(Q^2)$ correction to $S_{re}^<(K)$.

\section{Correlation functions with EM form factors}\label{sec_app_C}

The in-medium EM form factors modify the coupling between fermions and the external EM fields. Below we discuss how the modifications affect the calculation of correlation function in Sec.~\ref{sec_2}. The analysis depends crucially on expansion in momentum $Q$.
At $O(Q^0)$, the form factor $F_2$ is irrelevant. Using \eqref{spinor_products}, we easily find
\begin{align}
u^\dg(K_1)\G^\m u(K_2)A_\m=u^\dg(K_1)\s^\m u(K_2)A^F_\m+O(Q),
\end{align}
with $A_F^\m=(F_0 A_0,F_1 \vec{A})$. The effect of form factors at $O(Q^0)$ can be absorbed as rescaling factor to the gauge fields. The fact that $F_0\ne F_1$ is due to breaking of Lorentz symmetry by the medium. With the rescaled EM fields, we can write down the effective Lagrangian
\begin{align}\label{Lag_F}
{\cal L}=\bar{\psi}(i\pd_\m-A^F_\m)\g^\m\psi.
\end{align}
It follows that we should modify the gauge link accordingly $U(x_2,x_1)=e^{-i\int_{x_1}^{x_2} dz^\m A^F_\m(z)}$.
We then arrive at the following resummed correlation function at $O(Q^0)$
\begin{align}
S_{ra}^\m(K)=S_{re}^{ra}(K+A_F)=\frac{iK\cdot\bar{\s}}{K^2+i\h}.
\end{align}
Note that $K=P-A_F$ is the kinetic momentum with medium effect taken into account through $A\to A_F$. The calculations of $S_{ra}^\m$ at $O(Q^0)$ parallels to those in Sec.\ref{sec_2}.

Turning to $O(Q)$ contribution to $S_{ra}$ and noting that $K_{1,2}$ are on-shell up to $O(Q^2)$, we easily see the effect of in-medium EM form factors is to replace $\s^\m\to\G^\m$ in \eqref{scat}:
\begin{align}\label{S_ra_EM form factors}
S_{ra}=&\frac{iu(K_1)}{K_1^2+i\h}\big[u^\dg(K_1) i\G^\m u(K_2)A_\m\big]\frac{iu^\dg(K_2)}{K_2^2+i\h}\nonumber\\
=&\frac{iK_1\cdot\bar{\s}}{K_1^2+i\h}(-i A\cdot\G)\frac{iK_2\cdot\bar{\s}}{K_2^2+i\h},
\end{align}
whose components $S_{ra}^\m$ can be evaluated using \eqref{trace}. Again $S_{ar}$ can be obtained by flipping the sign of $\h$.

To convert $S_{ra}$ to $S^<$, we need to consider how the KMS relation in the presence of in-medium EM form factors. At $O(Q^0)$, the effective Lagrangian is essentially a free theory with interaction effect present only in the modified relation between canonical and kinetic momenta. The free theory clearly should satisfy KMS relation.

At $O(Q)$, the derivations of the KMS relation closely follow those in Appendix~\ref{sec_app_B}. Recall that an essential property used in the case without medium interaction is that the KMS relation holds at $O(Q^0)$, which is also valid in the presence of medium interaction as we reasoned above. It follows that the counterpart of \eqref{resum_KMS} is modified with $A\cdot\s\to A\cdot\G$
\begin{align}\label{Sless_EM form factors}
S^<_{re}(K)
=-2\p\left[-\tilde{f}(k_1^0)\d(K_1^2)K_1\cdot\bar{\s}(-iA\cdot \G )\frac{iK_2\cdot \bar{\sigma }}{K_2^2+i \eta }-\frac{-iK_1\cdot \bar{\sigma }}{K_1^2-i \eta }(-iA\cdot \G )K_2\cdot\bar{\s}\tilde{f}(k_2^0)\d(K_2^2)\right]_{K_1\to K_2+Q}.
\end{align}
Comparing \eqref{S_ra_EM form factors} and \eqref{Sless_EM form factors}, we confirm the KMS relation at $O(Q)$.




\section{Data Availability Statement}
No data associated in the manuscript.

\bibliographystyle{unsrt}

\begin{thebibliography}{100}

	\bibitem{Liang:2004ph} Zuo-Tang Liang and Xin-Nian Wang.
	\newblock {Globally polarized quark-gluon plasma in non-central A+A
		collisions}.
	\newblock {\em Phys. Rev. Lett.}, 94:102301, 2005.
	\newblock [Erratum: Phys.Rev.Lett. 96, 039901 (2006)].

	\bibitem{Liang:2004xn} Zuo-Tang Liang and Xin-Nian Wang.
	\newblock {Spin alignment of vector mesons in non-central A+A collisions}.
	\newblock {\em Phys. Lett. B}, 629:20--26, 2005.

	\bibitem{STAR:2017ckg} L.~Adamczyk et~al.
	\newblock {Global $\Lambda$ hyperon polarization in nuclear collisions:
		evidence for the most vortical fluid}.
	\newblock {\em Nature}, 548:62--65, 2017.

	\bibitem{STAR:2019erd} Jaroslav Adam et~al.
	\newblock {Polarization of $\Lambda$ ($\bar{\Lambda}$) hyperons along the beam
		direction in Au+Au collisions at $\sqrt{s_{_{NN}}}$ = 200 GeV}.
	\newblock {\em Phys. Rev. Lett.}, 123(13):132301, 2019.

	\bibitem{Becattini:2013fla} F.~Becattini, V.~Chandra, L.~Del~Zanna, and E.~Grossi.
	\newblock {Relativistic distribution function for particles with spin at local
		thermodynamical equilibrium}.
	\newblock {\em Annals Phys.}, 338:32--49, 2013.

	\bibitem{Fang:2016vpj} Ren-hong Fang, Long-gang Pang, Qun Wang, and Xin-nian Wang.
	\newblock {Polarization of massive fermions in a vortical fluid}.
	\newblock {\em Phys. Rev. C}, 94(2):024904, 2016.

	\bibitem{Li:2017slc} Hui Li, Long-Gang Pang, Qun Wang, and Xiao-Liang Xia.
	\newblock {Global $\Lambda$ polarization in heavy-ion collisions from a
		transport model}.
	\newblock {\em Phys. Rev. C}, 96(5):054908, 2017.

	\bibitem{Liu:2019krs} Shuai Y.~F. Liu, Yifeng Sun, and Che~Ming Ko.
	\newblock {Spin Polarizations in a Covariant Angular-Momentum-Conserved Chiral
		Transport Model}.
	\newblock {\em Phys. Rev. Lett.}, 125(6):062301, 2020.

	\bibitem{Becattini:2017gcx} F.~Becattini and Iu. Karpenko.
	\newblock {Collective Longitudinal Polarization in Relativistic Heavy-Ion
		Collisions at Very High Energy}.
	\newblock {\em Phys. Rev. Lett.}, 120(1):012302, 2018.

	\bibitem{Wei:2018zfb} De-Xian Wei, Wei-Tian Deng, and Xu-Guang Huang.
	\newblock {Thermal vorticity and spin polarization in heavy-ion collisions}.
	\newblock {\em Phys. Rev. C}, 99(1):014905, 2019.

	\bibitem{Wu:2019eyi} Hong-Zhong Wu, Long-Gang Pang, Xu-Guang Huang, and Qun Wang.
	\newblock {Local spin polarization in high energy heavy ion collisions}.
	\newblock {\em Phys. Rev. Research.}, 1:033058, 2019.

	\bibitem{Fu:2020oxj} Baochi Fu, Kai Xu, Xu-Guang Huang, and Huichao Song.
	\newblock {Hydrodynamic study of hyperon spin polarization in relativistic
		heavy ion collisions}.
	\newblock {\em Phys. Rev. C}, 103(2):024903, 2021.

	\bibitem{Zhang:2019xya} Jun-jie Zhang, Ren-hong Fang, Qun Wang, and Xin-Nian Wang.
	\newblock {A microscopic description for polarization in particle scatterings}.
	\newblock {\em Phys. Rev. C}, 100(6):064904, 2019.

	\bibitem{Weickgenannt:2020aaf} Nora Weickgenannt, Enrico Speranza, Xin-li Sheng, Qun Wang, and Dirk~H.
	Rischke.
	\newblock {Generating Spin Polarization from Vorticity through Nonlocal
		Collisions}.
	\newblock {\em Phys. Rev. Lett.}, 127(5):052301, 2021.

	\bibitem{Gao:2012ix} Jian-Hua Gao, Zuo-Tang Liang, Shi Pu, Qun Wang, and Xin-Nian Wang.
        \newblock {Chiral Anomaly and Local Polarization Effect from Quantum Kinetic Approach}.
        \newblock {\em Phys. Rev. Lett.}, 109:232301, 2012.

	\bibitem{Hidaka:2017auj} Yoshimasa Hidaka, Shi Pu, and Di-Lun Yang.
	\newblock {Nonlinear Responses of Chiral Fluids from Kinetic Theory}.
	\newblock {\em Phys. Rev. D}, 97(1):016004, 2018.

	\bibitem{Liu:2021uhn} Shuai Y.~F. Liu and Yi~Yin.
	\newblock {Spin polarization induced by the hydrodynamic gradients}.
	\newblock {\em JHEP}, 07:188, 2021.

	\bibitem{Becattini:2021suc} F.~Becattini, M.~Buzzegoli, and A.~Palermo.
	\newblock {Spin-thermal shear coupling in a relativistic fluid}.
	\newblock {\em Phys. Lett. B}, 820:136519, 2021.

	\bibitem{Fu:2021pok} Baochi Fu, Shuai Y.~F. Liu, Longgang Pang, Huichao Song, and Yi~Yin.
	\newblock {Shear-Induced Spin Polarization in Heavy-Ion Collisions}.
	\newblock {\em Phys. Rev. Lett.}, 127(14):142301, 2021.

	\bibitem{Becattini:2021iol} F.~Becattini, M.~Buzzegoli, G.~Inghirami, I.~Karpenko, and A.~Palermo.
	\newblock {Local Polarization and Isothermal Local Equilibrium in Relativistic
		Heavy Ion Collisions}.
	\newblock {\em Phys. Rev. Lett.}, 127(27):272302, 2021.

	\bibitem{Yi:2021ryh} Cong Yi, Shi Pu, and Di-Lun Yang.
	\newblock {Reexamination of local spin polarization beyond global equilibrium
		in relativistic heavy ion collisions}.
	\newblock {\em Phys. Rev. C}, 104(6):064901, 2021.

	\bibitem{Lin:2022tma} Shu Lin and Ziyue Wang.
	\newblock {Shear induced polarization: collisional contributions}.
	\newblock {\em JHEP}, 12:030, 2022.

	\bibitem{Liu:2020dxg} Shuai Y.~F. Liu and Yi~Yin.
	\newblock {Spin Hall effect in heavy-ion collisions}.
	\newblock {\em Phys. Rev. D}, 104(5):054043, 2021.

	\bibitem{Mameda:2022ojk} Kazuya Mameda, Naoki Yamamoto, and Di-Lun Yang.
	\newblock {Photonic spin Hall effect from quantum kinetic theory in curved
		spacetime}.
	\newblock {\em Phys. Rev. D}, 105(9):096019, 2022.

	\bibitem{RevModPhys.87.1213} Jairo Sinova, Sergio~O. Valenzuela, J.~Wunderlich, C.~H. Back, and
	T.~Jungwirth.
	\newblock Spin hall effects.
	\newblock {\em Rev. Mod. Phys.}, 87:1213--1260, Oct 2015.

	\bibitem{PhysRevLett.95.226801} C.~L. Kane and E.~J. Mele.
	\newblock Quantum spin hall effect in graphene.
	\newblock {\em Phys. Rev. Lett.}, 95:226801, Nov 2005.

	\bibitem{Fu:2022oup} Baochi Fu, Longgang Pang, Huichao Song, and Yi~Yin.
	\newblock {Baryonic spin Hall effects in Au+Au collisions at $\sqrt{s_{NN}} =
		7.7-200$ GeV}.
	\newblock 7 2022.

	\bibitem{Gao:2019znl} Jian-Hua Gao and Zuo-Tang Liang.
	\newblock {Relativistic Quantum Kinetic Theory for Massive Fermions and Spin
		Effects}.
	\newblock {\em Phys. Rev. D}, 100(5):056021, 2019.

	\bibitem{Weickgenannt:2019dks} Nora Weickgenannt, Xin-Li Sheng, Enrico Speranza, Qun Wang, and Dirk~H.
	Rischke.
	\newblock {Kinetic theory for massive spin-1/2 particles from the
		Wigner-function formalism}.
	\newblock {\em Phys. Rev. D}, 100(5):056018, 2019.

	\bibitem{Hattori:2019ahi} Koichi Hattori, Yoshimasa Hidaka, and Di-Lun Yang.
	\newblock {Axial Kinetic Theory and Spin Transport for Fermions with Arbitrary
		Mass}.
	\newblock {\em Phys. Rev. D}, 100(9):096011, 2019.

	\bibitem{Liu:2020flb} Yu-Chen Liu, Kazuya Mameda, and Xu-Guang Huang.
	\newblock {Covariant Spin Kinetic Theory I: Collisionless Limit}.
	\newblock {\em Chin. Phys. C}, 44(9):094101, 2020.
	\newblock [Erratum: Chin.Phys.C 45, 089001 (2021)].

	\bibitem{Sheng:2020oqs} Xin-Li Sheng, Qun Wang, and Xu-Guang Huang.
	\newblock {Kinetic theory with spin: From massive to massless fermions}.
	\newblock {\em Phys. Rev. D}, 102(2):025019, 2020.

	\bibitem{Yang:2020hri} Di-Lun Yang, Koichi Hattori, and Yoshimasa Hidaka.
	\newblock {Effective quantum kinetic theory for spin transport of fermions with
		collsional effects}.
	\newblock {\em JHEP}, 07:070, 2020.

	\bibitem{Weickgenannt:2021cuo} Nora Weickgenannt, Enrico Speranza, Xin-li Sheng, Qun Wang, and Dirk~H.
	Rischke.
	\newblock {Derivation of the nonlocal collision term in the relativistic
		Boltzmann equation for massive spin-1/2 particles from quantum field theory}.
	\newblock {\em Phys. Rev. D}, 104(1):016022, 2021.

	\bibitem{Sheng:2021kfc} Xin-Li Sheng, Nora Weickgenannt, Enrico Speranza, Dirk~H. Rischke, and Qun
	Wang.
	\newblock {From Kadanoff-Baym to Boltzmann equations for massive spin-1/2
		fermions}.
	\newblock {\em Phys. Rev. D}, 104(1):016029, 2021.

	\bibitem{Lin:2021mvw} Shu Lin.
	\newblock {Quantum kinetic theory for quantum electrodynamics}.
	\newblock {\em Phys. Rev. D}, 105(7):076017, 2022.

	\bibitem{Hidaka:2022dmn} Yoshimasa Hidaka, Shi Pu, Qun Wang, and Di-Lun Yang.
	\newblock {Foundations and applications of quantum kinetic theory}.
	\newblock {\em Prog. Part. Nucl. Phys.}, 127:103989, 2022.

	\bibitem{Sheng:2017lfu} Xin-li Sheng, Dirk~H. Rischke, David Vasak, and Qun Wang.
	\newblock {Wigner functions for fermions in strong magnetic fields}.
	\newblock {\em Eur. Phys. J. A}, 54(2):21, 2018.

	\bibitem{Lin:2019fqo} Shu Lin and Lixin Yang.
	\newblock {Chiral kinetic theory from Landau level basis}.
	\newblock {\em Phys. Rev. D}, 101(3):034006, 2020.

	\bibitem{Hattori:2016lqx} Koichi Hattori, Shiyong Li, Daisuke Satow, and Ho-Ung Yee.
	\newblock {Longitudinal Conductivity in Strong Magnetic Field in Perturbative
		QCD: Complete Leading Order}.
	\newblock {\em Phys. Rev. D}, 95(7):076008, 2017.

	\bibitem{Fukushima:2019ugr} Kenji Fukushima and Yoshimasa Hidaka.
	\newblock {Resummation for the Field-theoretical Derivation of the Negative
		Magnetoresistance}.
	\newblock {\em JHEP}, 04:162, 2020.

	\bibitem{Zhang:2020ben} Cheng Zhang, Ren-Hong Fang, Jian-Hua Gao, and De-Fu Hou.
	\newblock {Thermodynamics of chiral fermion system in a uniform magnetic
		field}.
	\newblock {\em Phys. Rev. D}, 102(5):056004, 2020.

	\bibitem{Fang:2021ndj} Ren-Hong Fang, Ren-Da Dong, De-Fu Hou, and Bao-Dong Sun.
	\newblock {Thermodynamics of the System of Massive Dirac Fermions in a Uniform
		Magnetic Field}.
	\newblock {\em Chin. Phys. Lett.}, 38(9):091201, 2021.

	\bibitem{Fukushima:2015wck} Kenji Fukushima, Koichi Hattori, Ho-Ung Yee, and Yi~Yin.
	\newblock {Heavy Quark Diffusion in Strong Magnetic Fields at Weak Coupling and
		Implications for Elliptic Flow}.
	\newblock {\em Phys. Rev. D}, 93(7):074028, 2016.

	\bibitem{Fukushima:2017lvb} Kenji Fukushima and Yoshimasa Hidaka.
	\newblock {Electric conductivity of hot and dense quark matter in a magnetic
		field with Landau level resummation via kinetic equations}.
	\newblock {\em Phys. Rev. Lett.}, 120(16):162301, 2018.

	\bibitem{Hattori:2017qih} Koichi Hattori, Xu-Guang Huang, Dirk~H. Rischke, and Daisuke Satow.
	\newblock {Bulk Viscosity of Quark-Gluon Plasma in Strong Magnetic Fields}.
	\newblock {\em Phys. Rev. D}, 96(9):094009, 2017.

	\bibitem{Hattori:2016njk} Koichi Hattori and Yi~Yin.
	\newblock {Charge redistribution from anomalous magnetovorticity coupling}.
	\newblock {\em Phys. Rev. Lett.}, 117(15):152002, 2016.

	\bibitem{Lin:2021sjw} Shu Lin and Lixin Yang.
	\newblock {Magneto-vortical effect in strong magnetic field}.
	\newblock {\em JHEP}, 06:054, 2021.

	\bibitem{Peng:2023rjj} Hao-Hao Peng, Xin-Li Sheng, Shi Pu, and Qun Wang.
	\newblock {Electric and magnetic conductivities in magnetized fermion systems}.
	\newblock {\em Phys. Rev. D}, 107(11):116006, 2023.

	\bibitem{Hattori:2023egw} Koichi Hattori, Kazunori Itakura, and Sho Ozaki.
	\newblock {Strong-Field Physics in QED and QCD: From Fundamentals to
		Applications}.
	\newblock 5 2023.

	\bibitem{Giunti:2014ixa} Carlo Giunti and Alexander Studenikin.
	\newblock {Neutrino electromagnetic interactions: a window to new physics}.
	\newblock {\em Rev. Mod. Phys.}, 87:531, 2015.

	\bibitem{Polyakov:2018zvc} Maxim~V. Polyakov and Peter Schweitzer.
	\newblock {Forces inside hadrons: pressure, surface tension, mechanical radius,
		and all that}.
	\newblock {\em Int. J. Mod. Phys. A}, 33(26):1830025, 2018.

	\bibitem{Buzzegoli:2021jeh} M.~Buzzegoli and Dmitri~E. Kharzeev.
	\newblock {Anomalous gravitomagnetic moment and nonuniversality of the axial
		vortical effect at finite temperature}.
	\newblock {\em Phys. Rev. D}, 103(11):116005, 2021.

	\bibitem{Lin:2023ass} Shu Lin and Tian Jia-Yuan.
	\newblock {Medium correction to gravitational form factors}.
	\newblock {\em Acta Phys. Sin.}, 72(7):071201, 2023.

	\bibitem{Hidaka:2016yjf} Yoshimasa Hidaka, Shi Pu, and Di-Lun Yang.
	\newblock {Relativistic Chiral Kinetic Theory from Quantum Field Theories}.
	\newblock {\em Phys. Rev. D}, 95(9):091901, 2017.

	\bibitem{le2000thermal} Michel Le~Bellac.
	\newblock {\em Thermal field theory}.
	\newblock Cambridge university press, 2000.
	

	\bibitem{Chou:1984es} K.~c.~Chou, Z.~b.~Su, B.~l.~Hao and L.~Yu,
	\newblock {Equilibrium and Nonequilibrium Formalisms Made Unified}.
	\newblock {\em Phys. Rept.}, 118:1 1985.
	

	\bibitem{Chen:2014cla} Jing-Yuan Chen, Dam~T. Son, Mikhail~A. Stephanov, Ho-Ung Yee, and Yi~Yin.
	\newblock {Lorentz Invariance in Chiral Kinetic Theory}.
	\newblock {\em Phys. Rev. Lett.}, 113(18):182302, 2014.

	\bibitem{Gagnon:2006hi} Jean-Sebastien Gagnon and Sangyong Jeon.
	\newblock {Leading order calculation of electric conductivity in hot quantum
		electrodynamics from diagrammatic methods}.
	\newblock {\em Phys. Rev. D}, 75:025014, 2007.
	\newblock [Erratum: Phys.Rev.D 76, 089902 (2007)].

	\bibitem{Gagnon:2007qt} Jean-Sebastien Gagnon and Sangyong Jeon.
	\newblock {Leading Order Calculation of Shear Viscosity in Hot Quantum
		Electrodynamics from Diagrammatic Methods}.
	\newblock {\em Phys. Rev. D}, 76:105019, 2007.

	\bibitem{Dong:2021fxn} Lihua Dong and Shu Lin.
	\newblock {Dilepton helical production in a vortical quark-gluon plasma}.
	\newblock {\em Eur. Phys. J. A}, 58(9):176, 2022.

	\bibitem{Becattini:2016gvu} F.~Becattini, I.~Karpenko, M.~Lisa, I.~Upsal and S.~Voloshin,
	\newblock {Global hyperon polarization at local thermodynamic equilibrium with vorticity, magnetic field and feed-down}.
	\newblock {Phys. Rev. C}, 95(5):054902 2017.


\end{thebibliography}


\end{CJK}

\end{document}